\documentclass[aps,pre,twocolumn,groupedaddress,showkeys]{revtex4-2}

\usepackage{graphicx} 
\usepackage{amsmath,amssymb,bm}
\usepackage{siunitx} 
\usepackage{hyperref} 
\usepackage{color}
\usepackage{orcidlink}

\begin{document}

\title{Moving Detector Quantum Walk with Random Relocation}
\author{Md Aquib Molla \orcidlink{0000-0003-0416-1349}}
\author{Sanchari Goswami \orcidlink{0000-0002-4222-5123}}
\affiliation{Vidyasagar College, Kolkata, India}
\date{\today}

\begin{abstract}
    We study a discrete-time quantum walk in presence of a detector at $x_D$ initially. The detector here is repeatedly removed after a span of $t_R$, the removal time, and reinserted at random locations. Two relocation rules are considered here: In Model~1, the detector is reinserted at any site beyond $x_D$, while in Model~2, reinsertion is done within a restricted window around the position of the detector at that time. Both variants behave like Semi Infinite Walk (SIW) for large $t_R$, where the detector behaves effectively as a fixed boundary. However, in the rapid-relocation regime, i.e., when $t_R$ is small, the behaviours are different. Model~1 permits greater spreading due to unrestricted reinsertion, which is different from Model~2. The time evolution of occupation probability ratio of our walker to that of an infinite walker at $x_D$, i.e., $f(x_D,t)/f_\infty(x_D,t)$, initially show the feature of a SIW upto $t=t_R$, then show some oscillatory behaviour and finally reach a saturation value for both the models. The ratio enhancing under certain conditions of $x_D$ and $t_R$, is a purely quantum mechanical effect. The saturation ratio shows a crossover behavior below and above a removal time $t_R^*$. At sites $x \neq x_D$ the occupation probablity ratios at a certain time reveals that for small $t_R$, the behaviours of the two models are drastically different from each other, as well as from Semi Infinite Walk (SIW), Quenched Quantum Walk (QQW) and Moving Detector Quantum Walk (MDQW). The correlation ratios of the two models with that of Infinite Walk (IW) show interesting time dependence for sites to the left or right of the initial detector position $x_D$.
\end{abstract}


\maketitle

\section{Introduction}

    Discrete-time quantum walks (DTQW) serve as a fundamental tool for quantum computation and algorithms. They are also used to study various physical systems and therefore to control and explain their dynamics \cite{Chandrasekar}. A few such examples are energy transport in photosynthesis, simulation of the Dirac equation, quantum magnetometry etc. \cite{Mohseni, Engel, Chandrasekhar_dirac, Razzoli}. The term Quantum Walk (QW) was first coined by Aharonov \textit{et al.}~\cite{Aharonov1993} . In contrast to classical  random walks (CRW), which is in, general diffusive, the interference between left- and right-moving amplitudes in quantum walk gives $<x^2> \sim t^2$~\cite{Nayak2001, Kempe2003, Ambainis2001}. The results from the continuous-time quantum walk and the discrete-time quantum walk are often similar, but due to the coin degree of freedom, the discrete-time variant has been shown to be more powerful than the other in some context\cite{Chandrasekhar_2}.

    Unlike the probability distribution of a classical walker, which is Gaussian, the probability distribution of a quantum walker is peaked at $x=\pm \frac{t}{\sqrt{2}}$. However, introducing a detector or absorbing boundary into the path of the walker~\cite{Bach2004, Kuklinski2020} significantly alters the probability distribution.  A stationary detector at site $x_D$ obstructs the walker to move to the right, producing a Semi-Infinite Walk (SIW). The corresponding probability distribution and other relevant quantities are studied in~\cite{Goswami2010}. When such a detector is removed after a finite time, the resulting walk is a quenched quantum walk (QQW)~\cite{Goswami2012}. This walk exhibits nontrivial enhancement of the occupation probability at $x_D$, along with characteristic scaling laws that depend on the removal time. These studies reveal that the scaling behaviour of different quantities for a QW are sensitive to measurement-induced boundaries.

    In this work, we study a QW with a detector which can detect upto a certain fixed time. After that time the detector hops to another site. The hop can also be thought of removal of the old detector and inserting a new one at some other site. The situation is extremely important in connection to experimental studies. Photonic quantum walk experiments, as mentioned in~\cite{Schreiber2010, Schreiber2011}, have shown that detector dead-time, finite efficiency, and reset operations can influence the probability distribution profiles of the walker. Practical setups, therefore, may require replacement or reinsertion of the detector during the experiment. This motivates the theoretical models where detector dynamics plays a vital role.

    A deterministic model of a moving detector was introduced in Ref.~\cite{Molla2023}, where the detector hops after a fixed number of detections. The study revealed important scaling behaviours of a few relevant quantities. The limiting behaviours of the walk as the infinite walk (IW), the SIW, and the QQW are observed under certain conditions involving the controlling parameters of the detector. In this work, we consider detector which hops to a random position. The models and measurement schemes are described in sections \ref{SecII} and \ref{SecIII} respectively. The notation summary is given in section \ref{SecIV} for better readability. The results are presented in section \ref{SecV}. Finally, discussions are made in section \ref{SecVI}.

\section{Model Description}\label{SecII}

    We introduce the \textit{Random-Relocation Moving-Detector Quantum Walk} (RR-MDQW), where the detector remains at $x_D$ for a time $t_R$ and is then removed. It is then relocated at a new position according to a stochastic rule. We consider two variants: 
    \begin{itemize}
        \item Model~1, where the detector is relocated at an arbitrarily chosen site strictly beyond $x_D$,
        \item Model~2, where the relocation is restricted between the detector position $x$ and $x+t_R$.
    \end{itemize}
    The two models differ in a sense that for Model~1, the direction of hop of the detector is not always fixed, whereas, for Model~2 it is always going towards the right. The common feature is that the detector begins at the site $x_D$ and remains perfectly absorbing whenever present. The details of the relocation schemes are presented in \ref{Mod1} and \ref{Mod2}. 
    
    \subsection{Model 1: Random Relocation Beyond $x_D$}\label{Mod1}

        In this model, the detector is placed at the initial position $x_D$ and remains there up to the removal time $t_R$. Once $t = t_R$ is reached, the detector is removed from $x_D$ and inserted at a randomly chosen lattice site strictly to the right of $x_D$, the initial position of the detector. The same procedure is repeated at every subsequent interval of duration $t_R$: at $t = 2t_R$, $3t_R$, and so on, the detector is again removed from its previous position and relocated at a new random position beyond $x_D$. Thus, the detector repeatedly reappears at arbitrary locations on the positive side of the lattice. It is worth mentioning that there is no upper bound on its displacement in this case.

    \subsection{Model 2: Random Relocation Within a Restricted Window}\label{Mod2}
        For the second model, the initial position of the detector $x_D$ and the removal time $t_R$ are both same as in Model~1. The difference arises after removal of the detector. Here, instead of placing the detector arbitrarily far to the right, we restrict its new position to a window within a further $t_R$.

        More precisely, after removal at time $t_R$, the detector is inserted at a site chosen uniformly at random from the interval
        \[
        x_D(\text{old}) \;\leqslant\; x_D(\text{new}) \;\leqslant\; x_D(\text{old}) + t_R.
        \]
        At the next removal time $t = 2t_R$, the same procedure is repeated, now using the updated detector location as the left boundary of the insertion interval. This process continues indefinitely.

        In this variant, the detector possesses an effective average velocity directed towards the positive side of the $x$-axis, since its admissible relocation window shifts rightward after each interval. The resulting motion shares similarities with the deterministic moving-detector model studied in Ref.~\cite{Molla2023}, although here the motion is constrained and stochastic rather than deterministic.

    \section{Measurement Scheme and Time Evolution of the Walk}\label{SecIII}

        In both variants of the Random-Relocation Moving-Detector Quantum Walk (RR-MDQW), the detector acts as a perfectly absorbing one whenever it is present. If the walker reaches the position of the detector $X_D(t)$ at time $t$, the corresponding probability amplitude is removed with unit probability ($p_D = 1$). Here we followed the approach of Ref.~\cite{Goswami2012}, where
        renormalization is deliberately avoided so that the true modification of occupation probabilities due to the detector dynamics can be captured.\\

        The state of the walker at position $x$ and time $t$ is represented by the two-component spinor
        \[
        \Psi(x,t)
        =
        \begin{pmatrix}
            \psi_L(x,t)\\[4pt]
            \psi_R(x,t)
        \end{pmatrix}
        \]
        where $L$ and $R$ denote left- and right-moving chirality states. The internal (coin) dynamics is governed by the Hadamard operator
        \[
        H = \frac{1}{\sqrt{2}}
        \begin{pmatrix}
            1 & 1 \\
            1 & -1
        \end{pmatrix}
        \]
        while the shift operator acts as
        \[
        T|x,L\rangle = |x-1,L\rangle, 
        \qquad  
        T|x,R\rangle = |x+1,R\rangle.
        \]
The full unitary update of the walker is
\[
\Psi(x,t+1) = T\, H\, \Psi(x,t),
\]
except at the detector position. 
If the detector is located at $X_D(t)$ at time $t$, then
\[
\Psi(X_D(t),t+1) = 0,
\]
reflecting perfect absorption. 
The occupation probability is given by
\[
f(x,t) = |\psi_L(x,t)|^2 + |\psi_R(x,t)|^2 .
\]

\section{Notation Summary} \label{SecIV}

For clarity, we summarize the notation used throughout the work:
\begin{itemize}
    \item $f(x,t)$: occupation probability of the RR-MDQW.
    \item $f_{\infty}(x,t)$: occupation probability of the corresponding infinite walk (IW).
    \item $S(t)$: survival probability of the walker at time $t$.
    \item $x_D$: initial position of the detector.
    \item $t_R$: time after which the detector is removed and relocated.
    \item $X_D(t)$: instantaneous position of the detector at time $t$.
    \item $(f/f_{\infty})_{\mathrm{sat}}$: long-time saturation value of the occupation probability ratio.
    \item Model~1: detector is relocated at any random site strictly to the right of $x_D$.
    \item Model~2: detector is relocated randomly within a restricted window 
    $[X_D(t),\, X_D(t)+t_R]$ after each removal.
\end{itemize}

 \begin{figure*}[t]
        \centering
        \includegraphics[width=0.5\textwidth,angle=-90]{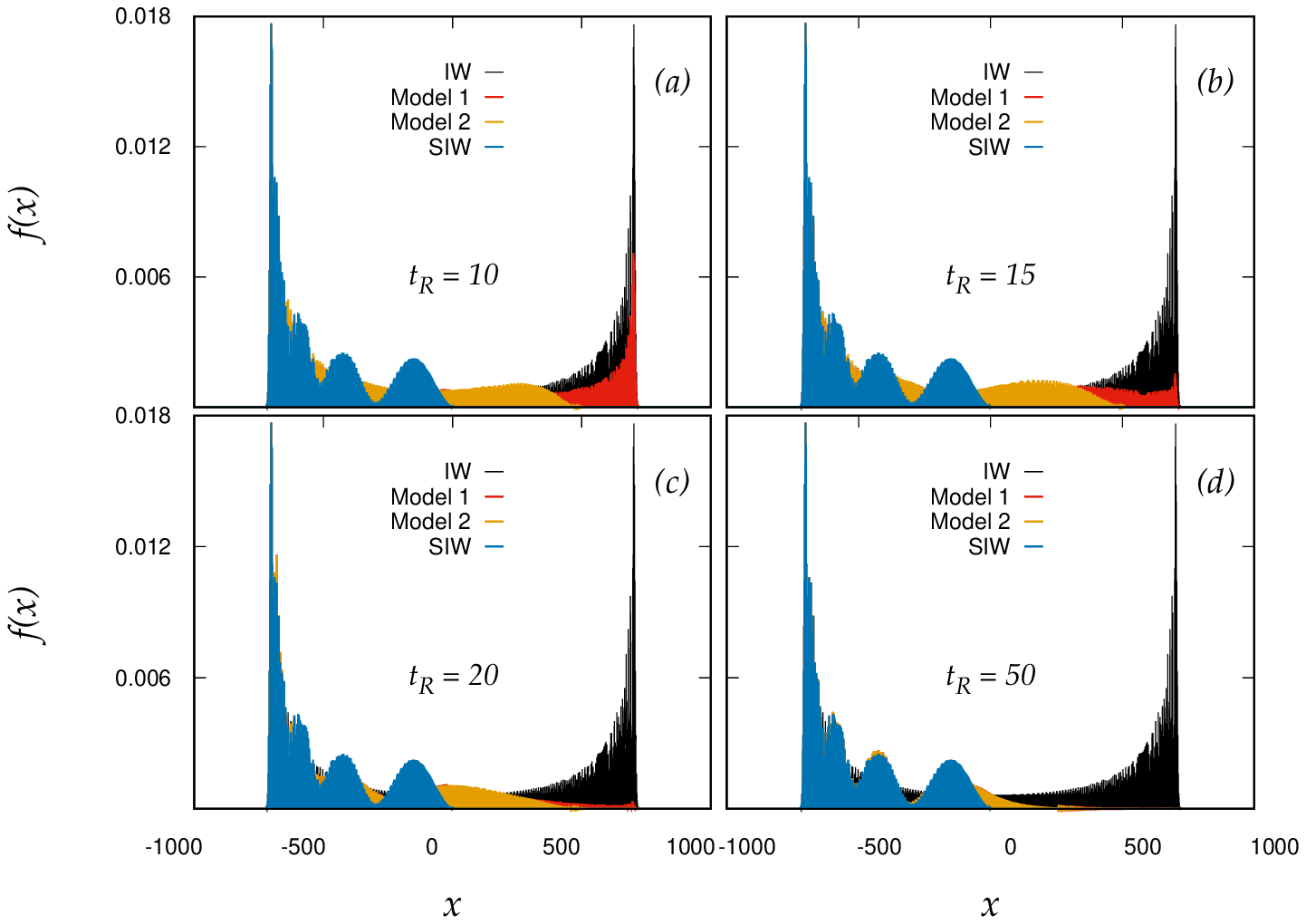}
        \caption{Comparison of the probability distributions for IW, SIW, Model~1, and Model~2 for different detector removal times : $(a)$ $t_R=10$, $(b)$ $t_R=15$, $(c)$ $t_R=20$ and $(d)$ $t_R=50$. In all the cases, initial position of the detector $x_D=10$. For large $t_R$ both models approach SIW, whereas for small $t_R$ they differ markedly: Model~1 allows wider spreading due to unrestricted relocation, while Model~2 remains more confined because of its bounded relocation window.}
        \label{fig:fig1}
    \end{figure*}

\section{Results}\label{SecV}

    \subsection{Probability distribution}

        We have shown the probability distributions of IW (black), Model 1 (red), Model 2 (orange) and SIW (blue) for $t = 1000$ in Fig. \ref{fig:fig1}. It has been observed that for large $t_R$ both Model 1 and Model 2 approach the SIW curve. The situation is however not the same for small $t_R$. For both the models the snapshots not only differ from SIW, rather they are different from each other also. For  $t_R$, small compared to $x_D$, Model 1 approaches to IW as it has the liberty to hop anywhere beyond $x_D$. On the other hand, in Model 2, due to the restriction imposed on the detector, that is it can only hop between $[X_D, X_D+t_R]$ where $X_D$ is the position of the detector at that time, the snapshot is different from IW. Although here the detector is always present, it is not the same as SIW. It is, at the same time, in no way similar to QQW or MDQW for small $t_R$.

        It is noted that in case of Model 2, for $t_R = 1$ the jumping window is too narrow, so that it can never actually jump from its initial position $x_D$. Therefore the resulting walk is a SIW. Thus, in Model 2, we get SIW in both the small and large limit of $t_R$. Also, in case of Model 1, if we increase the system size. $L \to \infty$, then upto time $t_R(\geqslant x_D)$ the walker gets detected $(t_R - x_D)/2$ times. It then hops to infinity and never comes back in the detection range. This occurs in case of QQW. Therefore, in the limit $L \to \infty$, Model 1 behaves like QQW.
        
        In the following subsections, we will try to compare the features of model~1 and model~2 to the IW in detail.

    \subsection{Probability ratio at $x_D$}
        
\begin{figure}[h!]
        \centering
        \includegraphics[width=0.75\linewidth,angle=-90]{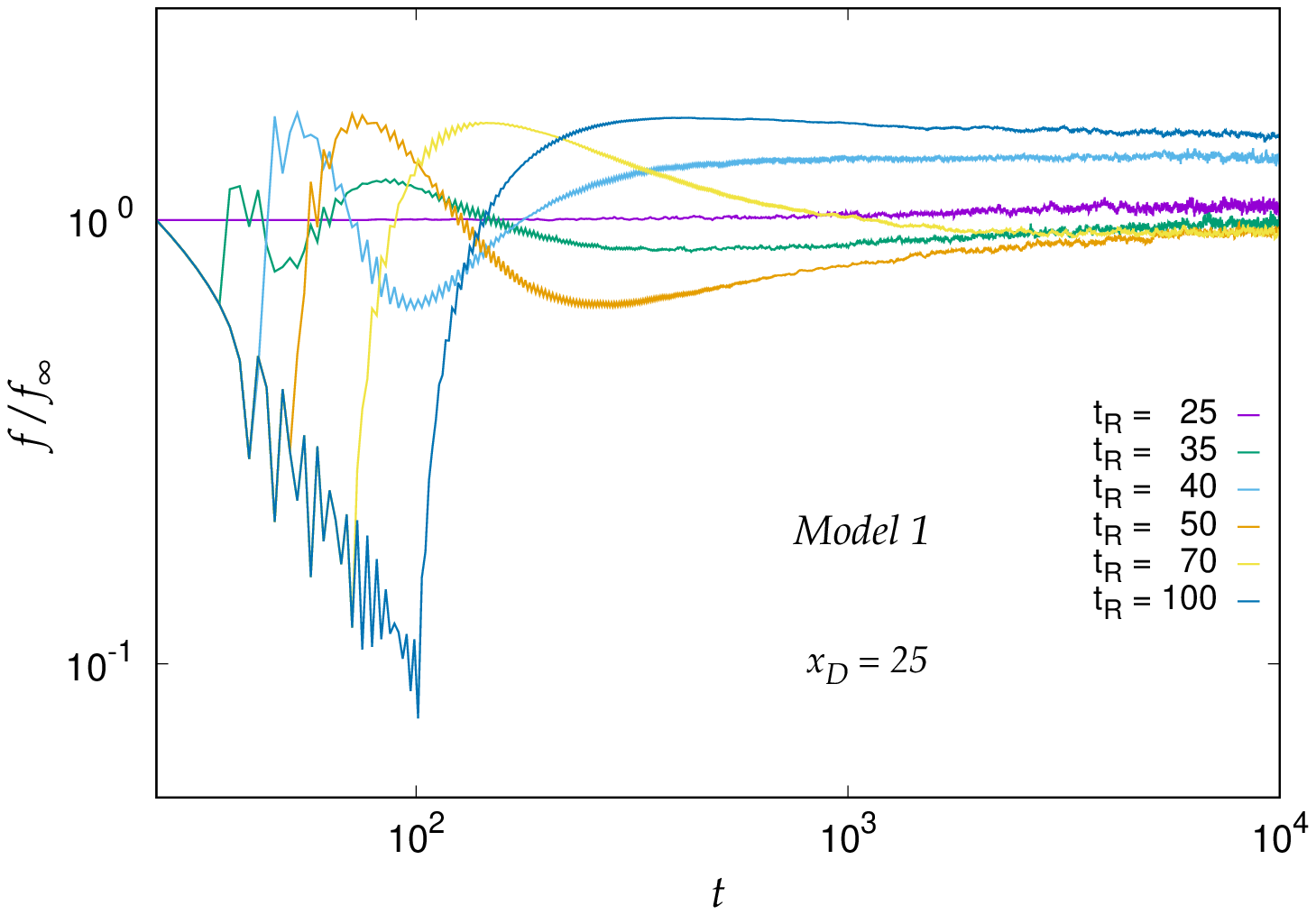}
        \caption{Ratio of the occupation probabilities of RR-MDQW to that of an Infinite Walk  \( f/f_{\infty} \) against time \(t\) for model~1, for \(x_D = 25\).}
        \label{fig:Model1_ff0_vs_t}
    \end{figure}

\begin{figure}[h!]
        \centering
        \includegraphics[width=0.75\linewidth,angle=-90]{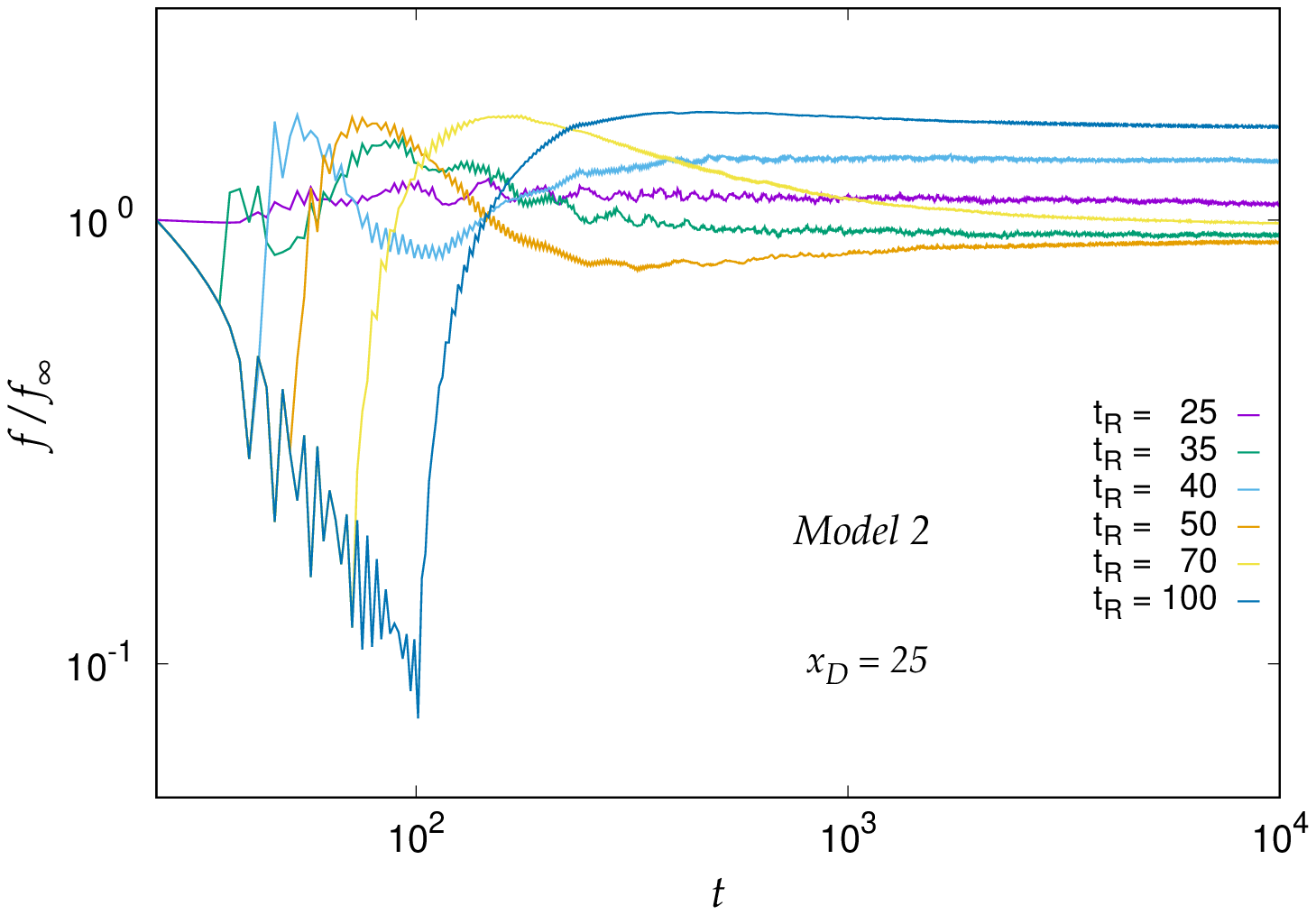}
        \caption{Ratio of the occupation probabilities of RR-MDQW to that of an Infinite Walk  \( f/f_{\infty} \) against time \(t\) for model~2, for \(x_D = 25\).}
        \label{fig:Model2_ff0_vs_t}
    \end{figure}

        We first compare the occupation probabilities of the walker at $x_D$ for different fixed removal times $t_R$ with that of IW. The ratio for $x_D$ at time $t$ may be written as $f(x_D,t)/f_{\infty}(x_D,t)$, or in shorthand notation as $f/f_{\infty}$. Figures~\ref{fig:Model1_ff0_vs_t} and~\ref{fig:Model2_ff0_vs_t} show the ratio $f/f_{\infty}$ as a function of $t$ for different choices of $t_R$ in Model~1 and Model~2. For small $t_R \sim x_D$, Model~1 behaves almost in the same way as the IW, and therefore the ratio stays close to unity throughout. This was also clear from the snapshot Fig. \ref*{fig:fig1}. The behaviour in Model~2 is noticeably different. Here, for low $t_R$ the detector repeatedly re-enters the path of the walker. Its frequent return and removal makes the walker prone to drift towards the positive side, where the detector is located. This means that the occupation probability at $x_D$ is enhanced, causing the ratio $f/f_{\infty}$ to saturate above unity. This increase is a genuinely quantum effect and is consistent with the observations reported for QQW and MDQW in Refs.~\cite{Goswami2012, Molla2023}. 
        
        However, for $t_R$ comparable to $x_D$ but larger, if we observe closely, for both the models $f/f_{\infty}$ show some astonishing behavior when observed with time. In the beginning, for any $t_R$, both of the models follow the SIW behaviour up to the time $t = t_R$, exhibiting a monotonic decay with time. As soon as the detector is removed, the ratio starts to rise. If $t_R$ is not too large compared to $x_D$, the ratio oscillates below and above unity before it reaches a saturation value $(f/f_{\infty})_{sat}$. As we increase $t_R$, the number of unity crossings (can be identified as IW) decrease. For higher but finite $t_R$, this ratio does not oscillate above and below unity, but, still saturates to a certain value $(f/f_{\infty})_{sat}$ corresponding to that $t_R$, which we will discuss later. For the case, as in Fig. \ref{fig:Model1_ff0_vs_t}, it has been observed that beyond $t_R>t_{cross}$, the ratio does not cross unity. In the limit $t_R \to \infty$, both models converge to the SIW.

    \begin{figure}[h!]
        \centering
        \includegraphics[width=0.7\linewidth,angle=-90]{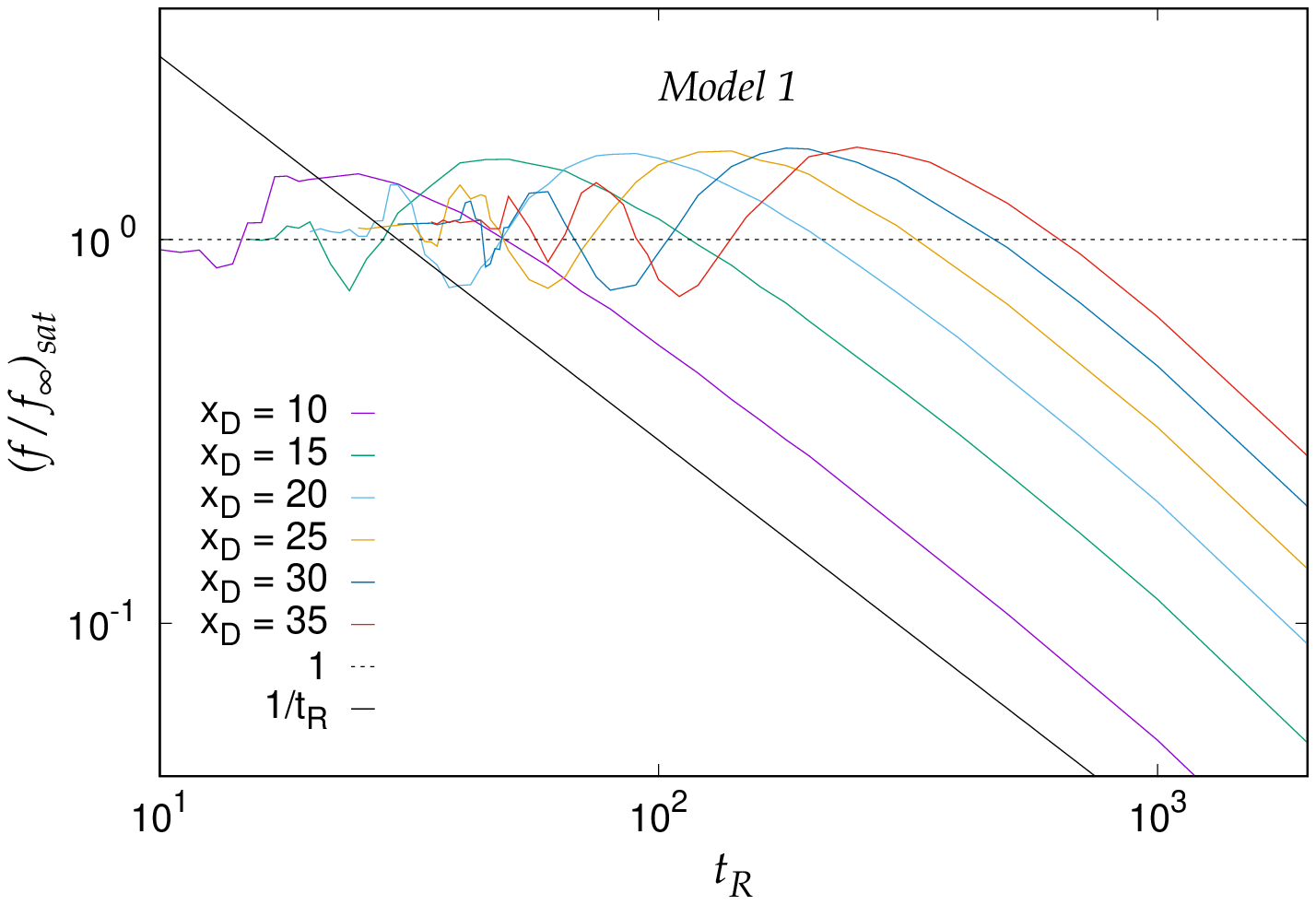}
        \caption{Variation of the saturation value of the occupation probability ratio \( (f/f_{\infty})_{sat} \) with detector removal time \(t_R\) for model~1, with $x_D$ as a parameter.}
        \label{fig:ff0Sat_vs_tR_Model1}
    \end{figure}

    \begin{figure}[h!]
        \centering
        \includegraphics[width=0.7\linewidth,angle=-90]{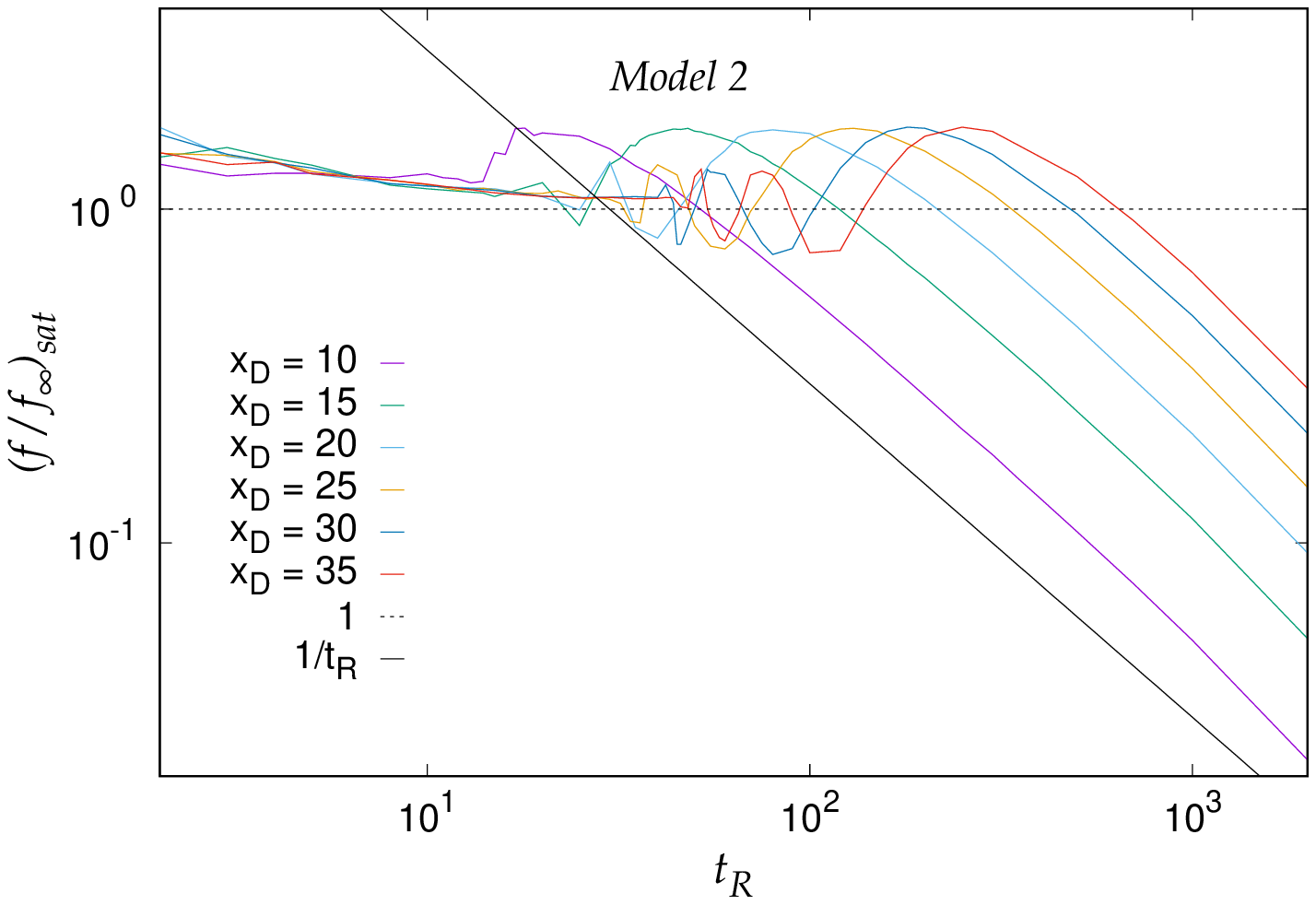}
        \caption{Variation of the saturation value of the occupation probability ratio \( (f/f_{\infty})_{sat} \) with detector removal time \(t_R\) for model~2, with $x_D$ as a parameter.}
        \label{fig:ff0Sat_vs_tR_Model2}
    \end{figure}
        
        In Figure \ref{fig:ff0Sat_vs_tR_Model1} and \ref{fig:ff0Sat_vs_tR_Model2}, we have shown the variations of \(\left( f/f_{\infty} \right)_{sat}\) against $t_R$. For both the models, \(\left( f/f_{\infty} \right)_{sat}\) shows an oscillatory behaviour for small $t_R$. 
                

    \begin{figure*}[t]
       \centering
        \includegraphics[width=0.5\textwidth,angle=-90]{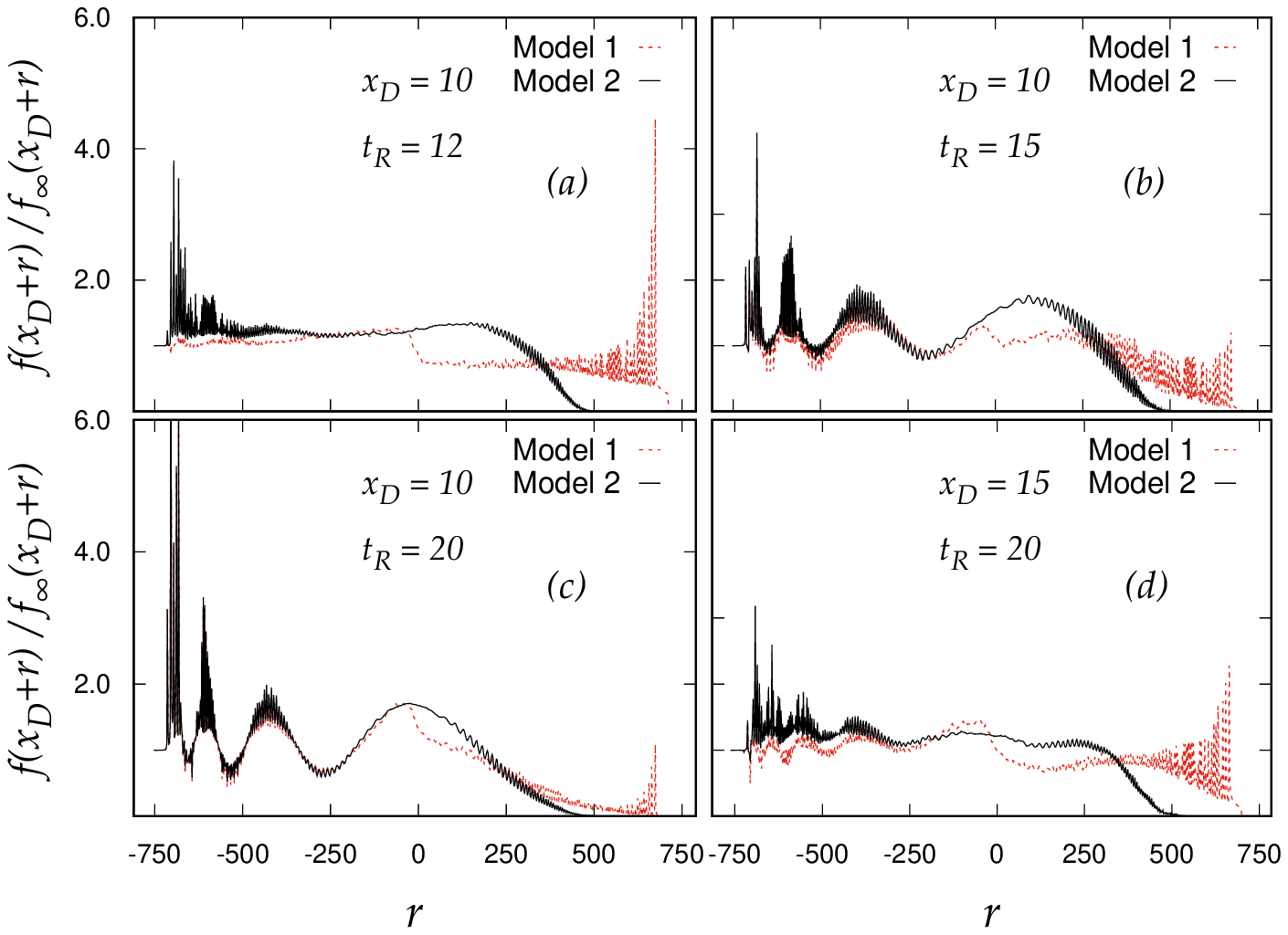}
        \caption{The ratio of the occupation probability distribution of RR-MDQW to that of the IW, $f(x_D+r)/f_{\infty}(x_D+r)$ as a function of $r$ with different combinations of $x_D$ and $t_R$. $(a)$ $x_D = 10$ and $t_R = 12$, $(b)$ $x_D = 10$ and $t_R = 15$, $(c)$ $x_D = 10$ and $t_R = 20$ and $(d)$ $x_D = 15$ and $t_R = 20$.}
        \label{fig:fig9_1000}
    \end{figure*}

    	For small values of $t_R$ (specifically when $t_R < x_D$), the detector is removed well before the walker typically reaches $x_D$. As a result, the first detection does not occur at $x_D$. Hence $f/f_{\infty}$ does not exhibit a saturation value in this regime for model~1.
        The situation is markedly different for model~2. Owing to the restriction on the motion of the detector, the walker remains confined within a finite interval. The occupation probability is enhanced at $x_D$ within this region. This in turn leads to a saturation value $\left(f/f_{\infty}\right)_{\mathrm{sat}}$ that lies above unity. For moderately small $t_R$, $\left(f/f_{\infty}\right)_{\mathrm{sat}}$ displays a weak oscillatory behaviour around unity, and we observe that the number of unity crossings increases with $x_D$ for both models. A further crossover appears beyond a characteristic time scale $t_R^{*}$. For $t_R > t_R^{*}$, the saturation value decays with $t_R$ according to
        \[
        \left( \frac{f}{f_{\infty}} \right)_{\mathrm{sat}} \bigg|_{t_R > t_R^{*}} \propto \frac{1}{t_R}.
        \]
        Moreover, we find that $t_R^{*}$ grows quadratically with the detector position, $t_R^{*} \propto x_D^{2}$. This scaling is consistent with the behaviour of the QQW reported in Ref.~\cite{Goswami2012}. It is worth mentioning that $t_{cross}$ behaves in a similar manner with $x_D$.
        
        The scaling behaviour of $\left( \frac{f}{f_{\infty}}\right)$ below and above $(t_R)^*$ is as follows:
        \begin{equation}\label{f_fsat}
        	\begin{aligned}
        		\left( \frac{f}{f_{\infty}}\right)_{sat} \sim \frac{1}{t_R}; \hspace{1.9 cm}t_R>(t_R)^*\\        		
        		\sim t_R \sin(1/t_R); \hspace{0.5cm}t_R<(t_R)^*,
        	\end{aligned}
        \end{equation}
      although for $t_R<t_R^*$, the behaviour is approximate.

        \begin{table}[h!]
		\centering
		\begin{tabular}{|c|cc|cc|}
			\hline
			 & \multicolumn{2}{c|}{No. of crossings} & \multicolumn{2}{c|}{$(t_R)_{\text{cross}}$} \\
			\cline{2-5}
			$x_D$ & Model 1 & Model 2 & Model 1 & Model 2 \\
			\hline
			$35$ & $5$ & $5$ & $700$ & $700$ \\
			\hline
			$30$ & $4$ & $4$ & $500$ & $500$ \\
			\hline
			$25$ & $4$ & $4$ & $300$ & $350$ \\
			\hline
			$20$ & $3$ & $3$ & $200$ & $200$ \\
			\hline
		\end{tabular}
		\caption{Approximate values of number and times of unity crossings as obtained from Figs.~4 and~5. The number of crossings represents how many times the curve crosses unity, and $(t_R)_{\text{cross}}$ gives the time beyond which the curve does not reach unity again.}
	\end{table}

    \begin{figure*}[t]
     	\centering
      	\includegraphics[width=0.5\textwidth,angle=-90]{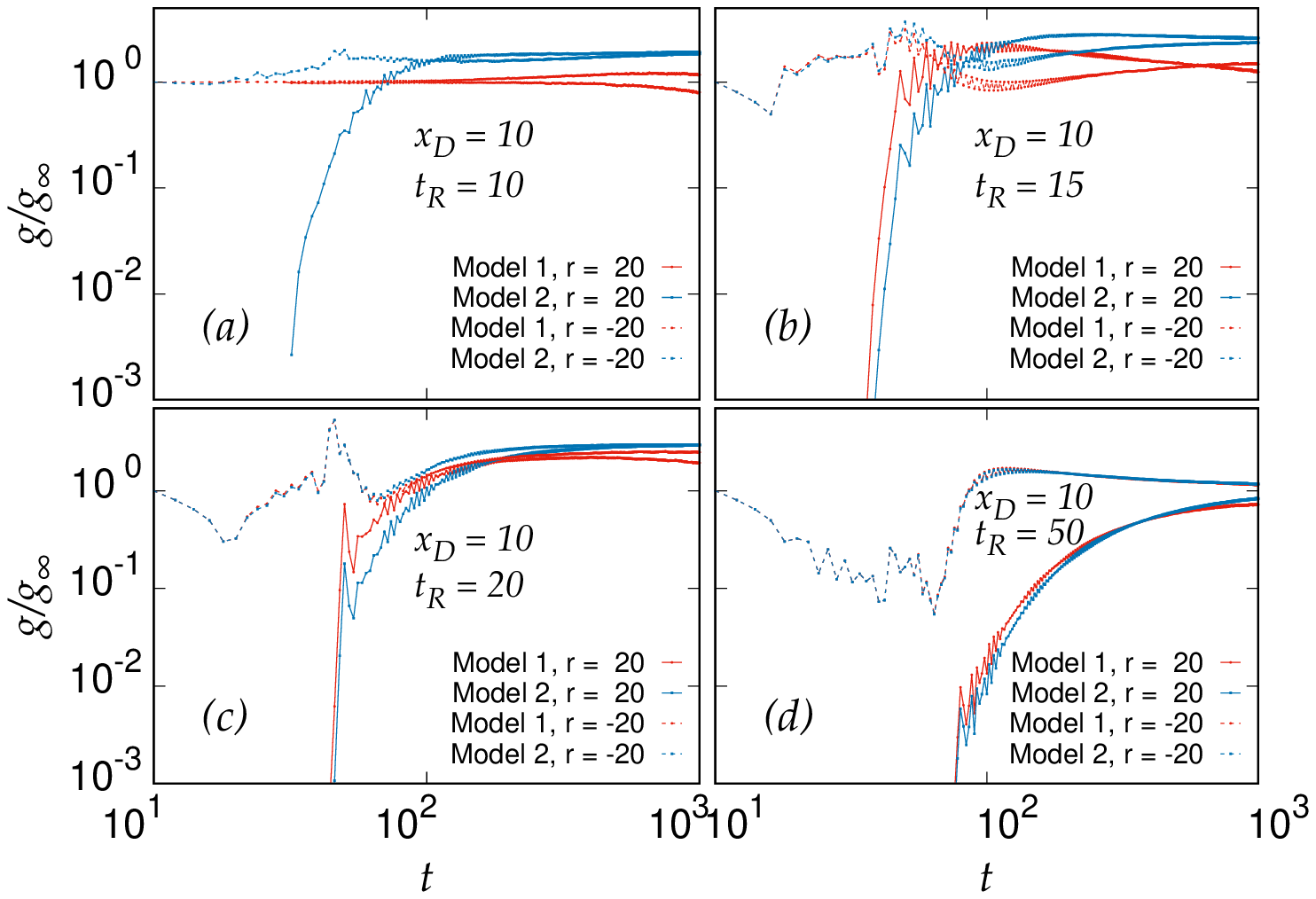}
      	\caption{The correlation ratio $g/g_{\infty}$ of Model~1 and Model~2 against $t$ with $x_D = 10$, $r = 20$ and $-20$ for different values of $t_R$ : $(a)$ $t_R = 10$, $(b)$ $t_R = 15$, $(c)$ $t_R = 20$ and $(d)$ $t_R = 50$.}
       	\label{correlation}
    \end{figure*}

\subsection{Probability ratio at a general $x$}

 So far, we have discussed the probability ratios at $x = x_D$. Now, we will discuss the probability ratios at sites $x$ which are not necessarily restricted to be $x_D$. Thus, a site, which is $r$ sites apart from $x_D$, can be described by the relation $x=x_D + r$, where $r$ can be positive or negative. From Fig. \ref{fig:fig9_1000}, we can observe the following :
 \begin{itemize}
    	\item For model~2, for $r<0$, several peaks are observed with the peak values much greater than $1$. For model~1 for large negative $r$, the ratio stays almost $1$ which implies that the RR-MDQW and IW behave in the same way in this region for model~1. The negative $r$ region gets affected more and more as $t_R$ is increased for both the models. It can be concluded here that memory effects for small $t_R$ and $r<0$ are strong for model~1 but not that much for model~2. 
       	\item For $r>0$, for large $t_R$, the two models behave in almost the same way; the ratios diminish for large $r$. For small $t_R$, the ratio for model~1 is much above $1$ as we go to high $r$, whereas for model~2 it gradually decreases with $r$. As $t_R$ is increased, the ratio toggles between $0$ (or a low value above $0$) and $1$ for model~1 and finally becomes $0$. The behaviour is different for model~2. Here the ratio falls off sharper without much oscillation.    
  \end{itemize}


\subsection{Correlations}\label{corr}

   To characterize the spatial dependence of the occupation probability away from the detector site, it is useful to examine correlations between different lattice positions. In particular, we focus on correlations between the detector site $x_D$ and a site displaced by a distance $r$ ($r$ can be both positive and negative). For the RR-MDQW, we define the equal-time correlation function as
    \begin{equation}
        \begin{aligned}
             g(x_D+r,t) &= f(x_D+r,t)\,f(x_D,t) \\
             g_{\infty}(x_D+r,t) &= f_{\infty}(x_D+r,t)\,f_{\infty}(x_D,t)
        \end{aligned}
    \end{equation}

The ratio of these two correlation functions $\left( \frac{g}{g_{\infty}} \right)$, provides a normalized measure of how detector relocation modifies spatial correlations relative to the IW. The correlation ratio $g/g_{\infty}$ depends strongly on whether $r<0$ or $r>0$, as well as on the detector removal time $t_R$.

    For sites located to the left of the detector ($r<0$), Fig.~\ref{correlation}(a) shows that for small $t_R$ the behaviour of Model~1 remains close to that of the IW. In contrast, Model~2 exhibits a markedly different behaviour. The correlation ratio here saturates well above unity, which is obviously due to the repeated removal and insertion of the detector in a narrow window. As $t_R$ is increased [Figs.~\ref{correlation}(b)--(d)], the distinction between the two models gradually diminishes. In both the models, the correlation ratios saturate above unity if $t_R$ is not much larger compared to $x_D$. In this regime, the detector remains at a given site for longer durations, and both models converge towards SIW-like behaviour. 

For sites to the right of the detector ($r>0$), a similar enhancement of the correlation ratio above unity is observed for Model~2 at small $t_R$, as shown in Fig.~\ref{correlation}(a). Model~1, on the other hand, behaves essentially like the IW, for reasons already discussed. With increasing $t_R$ [Figs.~\ref{correlation}(b) and (c)], the saturation value of the correlation ratio decreases progressively, while remaining above unity. For sufficiently large $t_R$, compared to $x_D$, [Fig.~\ref{correlation}(d)], the saturation value drops below unity.

It is worth noting that beyond $t_R \simeq 20$, the saturation value of the correlation ratios decrease systematically for both $r<0$ and $r>0$. In the large-$t_R$ regime, the correlation ratio approaches its saturation value from above for $r<0$ and from below for $r>0$, reflecting the asymmetric influence of the detector on the two sides of the lattice. It is also to be noted that beyond  $t_R \simeq 20$, the saturation values depend on the magnitude of $r$ and whether $r<0$ or $r>0$, and not on the specific model. Beyond this limit of $t_R$ corresponding to the $x_D$ value, the difference between the two models is lost. In the large $t_R$ limit, we can conclude that although the system try to approach the IW picture, it will never reach the same.  

\section{Summary and Discussions}\label{SecVI}

    In this work, we have studied the detailed effect of a detector for a quantum system. This has been done here for a quantum walker where the detector is placed in its path initially at a position and then removed and relocated at other positions. Two relocation schemes have been studied : 
    In Model~1, the detector is relocated arbitrarily far to the right for which the walker frequently encounters long intervals without any effective boundary, allowing significant spilling beyond $x_D$ and thereby producing a broader distribution closer to the IW. In contrast, Model~2 restricts the relocation window to a restricted interval that shifts slowly with time, thereby keeping the detector always relatively close to the walker. This restriction results in a more confined distribution with stronger suppression on the side where the detector is present. Thus, the two models represent different stochastic mechanisms. The statistical differences of Model~1 and Model~2 become more pronounced when relocations occur frequently. 
        The comparison among the IW, the SIW, and RR-MDQW (Model~1 and Model~2) have been presented in Figure~\ref{fig:fig1}. The IW shows symmetric ballistic spreading, while the SIW profile is truncated at the detector location. Both Model~1 and Model~2 lie between these two limiting cases under certain conditions of $x_D$ and $t_R$.

    As there are enormous number of QW experiments in recent years, the role and limitations of detectors is a very important subject to study. Like QQW, MDQW, here also it is evident that the occupation probability of sites may be enhanced compared to IW under certain conditions. This is a purely quantum mechanical effect. A detector with a very high efficiency can be thought of as a detector with high $t_R$. If such a detector is placed at a site towards right, then the occupation probability cannot approach the IW picture on the right, but the walker resembles a SIW picture. In our work, for large $t_R$, both the models approach SIW. For any moderate $t_R$, the time evolution of occupation probability ratio $f/f_\infty$, initially shows the feature of a SIW upto $t=t_R$. After that, there is some oscillatory behaviour and finally the ratio reaches a saturation value for both the models (Fig.~\ref{fig:Model1_ff0_vs_t} and \ref{fig:Model2_ff0_vs_t}). The ratio enhancing under certain conditions of $x_D$ and $t_R$, is a purely quantum mechanical effect. The saturation ratio $(f/f_{\infty})_{sat}$ shows a crossover behavior below and above a removal time $t_R^*$. Below $t_R^*$, the saturation ratio behaves approximately as $t_R \sin(1/t_R)$, whereas, above $t_R^*$, it behaves as $1/t_R$ (Fig.~\ref{fig:ff0Sat_vs_tR_Model1} and \ref{fig:ff0Sat_vs_tR_Model2}). 

For sites $x \neq x_D$, (a site $r$ apart from $x_D$ is $x_D+r$, r being both positive and negative). the two models are noticeably different in the small $t_R$ regime. This is clear from Fig.~\ref{fig:fig9_1000}, where $f(x_D+r)/f_{\infty}(x_D+r)$ is shown. In this regime, the models are not similar to QQW, MDQW. It has been observed here that memory effects for small $t_R$ and $r<0$ are strong for model~1 but not that much for model~2. For $r>0$ and for small $t_R$, the ratio for model~1 is much above $1$ for high $r$, whereas for model~2 it gradually decreases with $r$. When $t_R$ is small, the correlation ratios for model~1 and model~2 saturate to different values irrespective of whether $r<0$ or $r>0$. When $t_R$ is sufficiently large compared to $x_D$, the saturation values depend on whether $r<0$ or $r>0$, irrespective of the model. In any case, the models can never approach the IW picture for large $t_R$, which affect the system significantly.

The present work can be extended by studying the response of the system when $p_D$, the absorption probability of the detector is some definite function of time. Replacement of the detector at the same position but after a finite time span repeatedly can also be another interesting thing to study, which will be studied in near future. 

AM acknowledges financial support from CSIR, India (Grant no.
08/0463(12870)/2021-EMR-I). AM and SG acknowledge
the computational facility of Vidyasagar College, University of Calcutta.

\end{document}